\documentclass[a4paper,fleqn,usenatbib]{mnras}

\usepackage[T1]{fontenc}
\usepackage{ae,aecompl}

\usepackage{graphicx}	
\usepackage{amsmath}	
\usepackage{amssymb}	

\title[LRG-BEASTS III: WASP-80b]{LRG-BEASTS III: Ground-based transmission spectrum of the gas giant orbiting the cool dwarf WASP-80}

\author[J. Kirk et al.]{
J. Kirk,$^{1}$\thanks{E-mail: James.Kirk@warwick.ac.uk}
P. J. Wheatley\thanks{E-mail: P.J.Wheatley@warwick.ac.uk},$^{1}$
T. Louden,$^{1}$
I. Skillen,$^{2}$ 
G. W. King,$^{1}$
J. McCormac$^{1}$, \newauthor
and P. G. J. Irwin$^{3}$
\\
$^1$Department of Physics, University of Warwick, Coventry, CV4 7AL, UK \\
$^{2}$Isaac Newton Group of Telescopes, Apartado de Correos 321, 38700 Santa Cruz de Palma, Spain\\
$^{3}$Atmospheric, Oceanic, and Planetary Physics, Clarendon Laboratory, University of Oxford, Parks Road, Oxford OX1 3PU, UK\\
}

\date{Accepted XXX. Received YYY; in original form ZZZ}

\pubyear{2017}

\begin{document}
\label{firstpage}
\pagerange{\pageref{firstpage}--\pageref{lastpage}}
\maketitle

\begin{abstract}
We have performed ground-based transmission spectroscopy of the hot Jupiter orbiting the cool dwarf WASP-80 using the ACAM instrument on the William Herschel Telescope (WHT) as part of the LRG-BEASTS programme. This is the third paper of a ground-based transmission spectroscopy survey of hot Jupiters using low-resolution grism spectrographs. We observed two transits of the planet and have constructed transmission spectra spanning a wavelength range of 4640 -- 8840\,\AA. Our transmission spectrum is inconsistent with a previously claimed detection of potassium in WASP-80b's atmosphere, and is instead most consistent with a haze. We also do not see evidence for sodium absorption at a resolution of 100\,\AA.
\end{abstract}

\begin{keywords}
methods: observational -techniques: spectroscopic -planets and satellites: individual: WASP-80b -planets and satellites: atmospheres
\end{keywords}



\section{Introduction}

The Low Resolution Ground-Based Exoplanet Atmosphere Survey using Transmission Spectroscopy (LRG-BEASTS, `large beasts') is a programme to characterise hot Jupiter atmospheres homogeneously. This is the third planet in our survey, following the detection of a haze-induced Rayleigh scattering slope in the atmosphere of HAT-P-18b \citep{Kirk2017} and clouds in the atmosphere of WASP-52b \citep{Louden2017}. 

Transmission spectroscopy is revealing a diverse array of exoplanet atmospheres, from clear (e.g. WASP-39b: \citealt{Fischer2016}; \citealt{Nikolov2016}; \citealt{Sing2016}) to hazy (e.g. HAT-P-18b: \citealt{Kirk2017}) to cloudy (e.g. WASP-52b: \citealt{Kirk2016}; \citealt{Chen2017_w52}; \citealt{Louden2017}). As yet, no clear correlation has emerged between fundamental planetary parameters and the broad atmospheric characteristics (e.g. \citealt{Sing2016}), although there is tentative evidence that hotter exoplanets are more likely to be cloud free \citep{Heng2016}. This makes additional studies necessary to shed light on such correlations if they exist, and LRG-BEASTS, along with other ground-based surveys such as the Gran Telescopio Canarias (GTC) exoplanet transit spectroscopy survey (e.g. \citealt{Parviainen2016}, \citealt{Chen2017_w52}), ACCESS \citep{Rackham2017}, the Gemini/GMOS Transmission Spectral Survey \citep{Huitson2017}, and the VLT/FORS2 survey (e.g. \citealt{Nikolov2016}; \citealt{Gibson2017}) will expand the sample of studied planets. 

While low-resolution transmission spectroscopy can probe the deeper, pressure-broadened features in the atmosphere and reveal clouds and hazes, high-resolution transmission spectroscopy can reveal narrow line features at lower pressures. Recent high-resolution results include the measurements of the wind speed on HD\,189733b \citep{Louden2015,Brogi2016}, abnormal variability in the stellar line profiles of H$_\alpha$ near the transit of HD\,189733b \citep{Cauley2017}, detections of exoplanetary sodium (e.g. \citealt{Wyttenbach2015}; \citealt{Khalafinejad2017}; \citealt{Wyttenbach2017}), and analysing the affect of centre-to-limb variation on transmission spectra \citep{Yan2017}.

WASP-80b is a gas giant, with a radius of 0.986\,R$_{\textrm{J}}$, orbiting a cool dwarf, with a radius of 0.593\,R$_\odot$ \citep{Mancini2014_w80}. This puts it in a rare class of objects, and its large transit depth of 2.9\% makes it a good candidate for transmission spectroscopy. For this reason, there have been previous atmospheric studies of WASP-80b.

Transit photometry of WASP-80b has suggested a hazy atmosphere with no large variation with planetary radius (\citealt{Fukui2014}; \citealt{Mancini2014_w80}; \citealt{Triaud2015}). However, \cite{Sedaghati2017}, using the VLT/FORS2, reported a detection of pressure-broadened potassium absorption. This suggests a clear and low metallicity atmosphere, as clouds or hazes would act to mask the wings of this feature. In contrast, \cite{Parviainen2017} recently published a transmission spectrum of WASP-80b, using the GTC, that was best represented by a flat line and showed no evidence for potassium absorption.
 
In this paper, we present a low resolution transmission spectrum of WASP-80b with ACAM, as part of our LRG-BEASTS programme, which is inconsistent with the claimed potassium feature.

\section{Observations}

Two transits of WASP-80b were observed on the nights of 2016 August 18 and 2016 August 21 using the ACAM instrument \citep{Benn2008} on the William Herschel Telescope (WHT). This is the same instrument and setup we used in \cite{Kirk2017}. 

The observations were taken in fast readout mode with a smaller-than-standard window to reduce the overheads to 10\,s, with exposure times of 50\,s. For these transits we used a 40 arcsec wide slit as the 27 arcsec slit used in our study of HAT-P-18b was broken. We chose to use this wide slit to avoid differential slit losses between the target and comparison star. Due to the relatively sparse field, this wide slit did not cause problems with contaminating stars (Fig. \ref{fig:extraction_frame}, top panel). As with \cite{Kirk2017}, we chose not to use an order blocking filter due to concerns this may introduce unwanted systematics. Biases, flat fields and arc spectra were taken at the start and end of both nights. 

On the first night we observed the target from airmass 1.41 to 1.16 to 1.30, with a moon illumination of 99\,\% at 33 degrees from the target. On the second night we observed the target from airmass 1.20 to 1.16 to 1.70, with a moon illumination of 83\,\% at 73 degrees from the target. The second night was affected by clouds passing overhead before and during ingress but these cleared by the time of mid-transit (Fig. \ref{fig:ancillary_plots}).

To perform differential spectroscopy, we simultaneously observed a comparison target with a similar magnitude to that of WASP-80. The colour difference between the two stars was larger than desirable as we were limited by the length of the ACAM slit, which is 7.6 arcmin, and the fact that WASP-80 has a very close companion. This companion needed to be oriented such that its light was not blended with that of the target (Fig. \ref{fig:extraction_frame}). The chosen comparison star is 4.3 arcmin from the target and has a $V$ magnitude of 12.6 and $B-V$ colour of 0.97. This compared to WASP-80's $V$ magnitude of 11.9 and $B-V$ colour of 1.38.

\section{Data Reduction}

To reduce the data, we used the same custom \textsc{python} scripts as introduced in \cite{Kirk2017}. For the first night, 102 bias frames were median combined to create a master bias and for the second night, 111 bias frames were median combined. To flat-field our science images, we used a spectral sky flat normalised with a running median, taken at twilight with the same 40 arcsec wide slit that was used for our science images. To create the sky flat, 66 flat frames were median combined for night 1 and 86 flat frames for night 2. 

\begin{figure}
\centering
\includegraphics[scale=0.65]{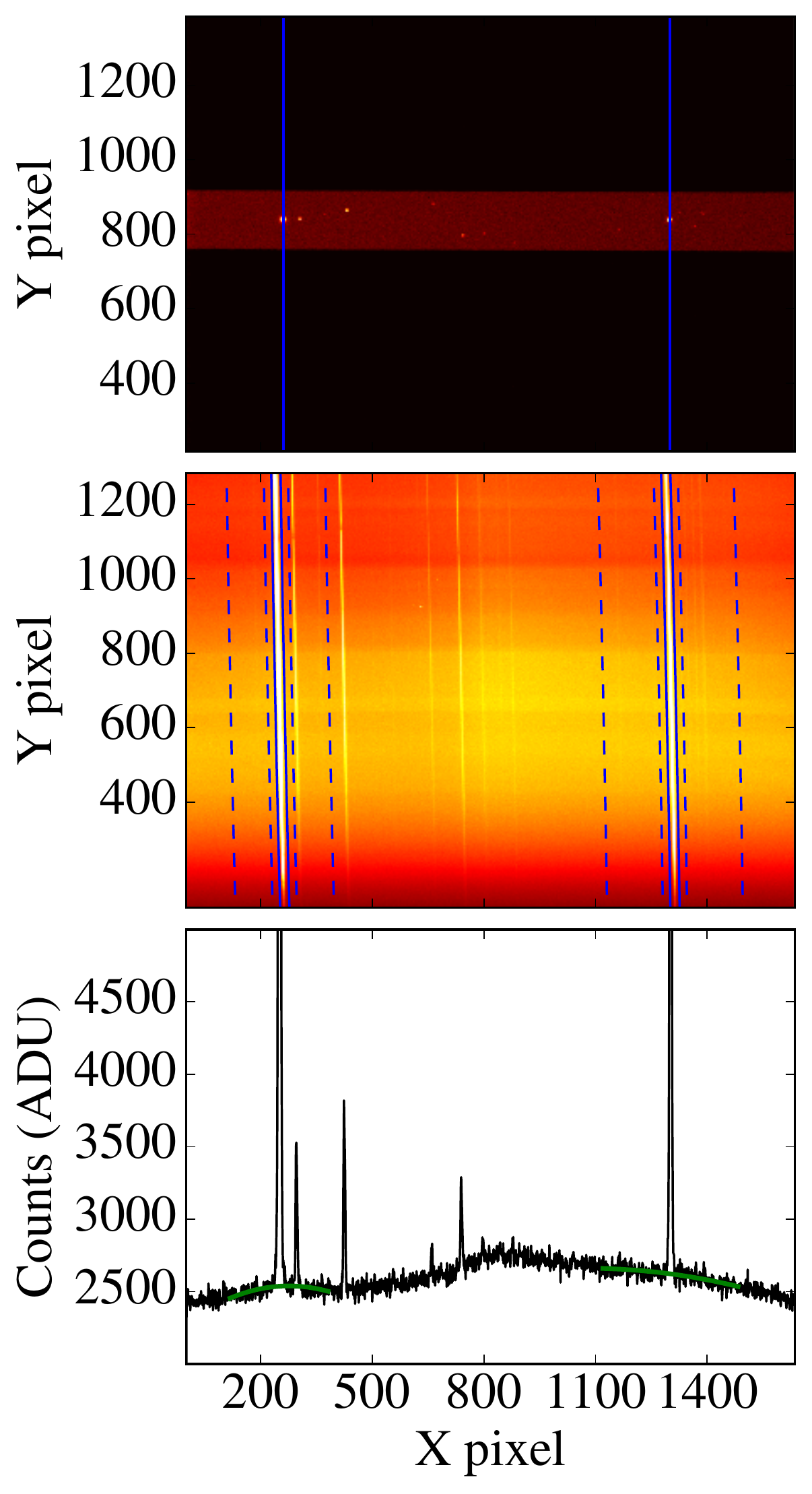}
\caption{Top panel: a through-slit image of the field, taken with a 40 arcsec wide slit. The target is the star intersected by the left vertical line and the comparison is the star intersected by the right vertical line. The image has been cropped in the vertical direction for clarity. Middle panel: Example science frame following bias and flat field corrections, with blue wavelengths at the bottom of the CCD and red wavelengths at the top. The extraction apertures are shown by the solid blue lines for the target (left) and comparison (right). The dashed lines indicate the regions in which the sky background was estimated. Contaminating stars falling within these regions were masked from the background fit. The close companion to the target as noted in the text can be seen as the spectral trace immediately to the right of the target trace. This was masked from the background estimation in this region. Bottom panel: A cut along the spatial direction at a y-pixel of 740. The background polynomial fits to each star are shown by the green lines. The y-axis has been cropped for clarity.}
\label{fig:extraction_frame}
\end{figure}

To extract the spectra of both stars, we followed the process as described in \cite{Kirk2017}. This involved fitting a polynomial to the locations of the traces, fitting the sky background across two regions either side of each trace, and performing normal extraction with a fixed aperture (Fig. \ref{fig:extraction_frame}). We experimented with the choice of extraction aperture width, background offset, background width and the order of the polynomial fitted across the background. In each case, we fitted an analytic transit light curve with a cubic polynomial to the resulting white light curve and used the extraction parameters that produced the lowest RMS in the residuals. We found that a 25 pixel-wide aperture produced the lowest scatter. The pixel scale of ACAM is 0.25 arcsec pixel$^{-1}$.

The background was estimated by fitting a quadratic polynomial across two regions either side of the target and comparison traces after masking contaminating stars from these regions (Fig. \ref{fig:extraction_frame}). For the target these regions were 100 pixels wide, and offset by 20 pixels from the target aperture. For the comparison these were again offset by 20 pixels from the comparison aperture but 150 pixels wide. The extra width was needed for the comparison as several contaminating stars had to be masked from these regions, while for the target a narrower region was used to avoid getting too close to the left hand edge of the chip (Fig. \ref{fig:extraction_frame}). We found that the combination of these background widths with the quadratic polynomial modelled the background variation well (Fig. \ref{fig:extraction_frame}). Cosmic rays were removed from the background regions by masking pixels that deviated by greater than $3\sigma$ from the median. Cosmic rays falling within the target aperture were removed in the same way as described in \cite{Kirk2017}, by dividing each spectrum by a frame clean of cosmics and removing deviant points. These reductions resulted in 230 spectra for night 1 and 237 spectra for night 2. 

\begin{figure*}
\centering
\includegraphics[scale=0.3]{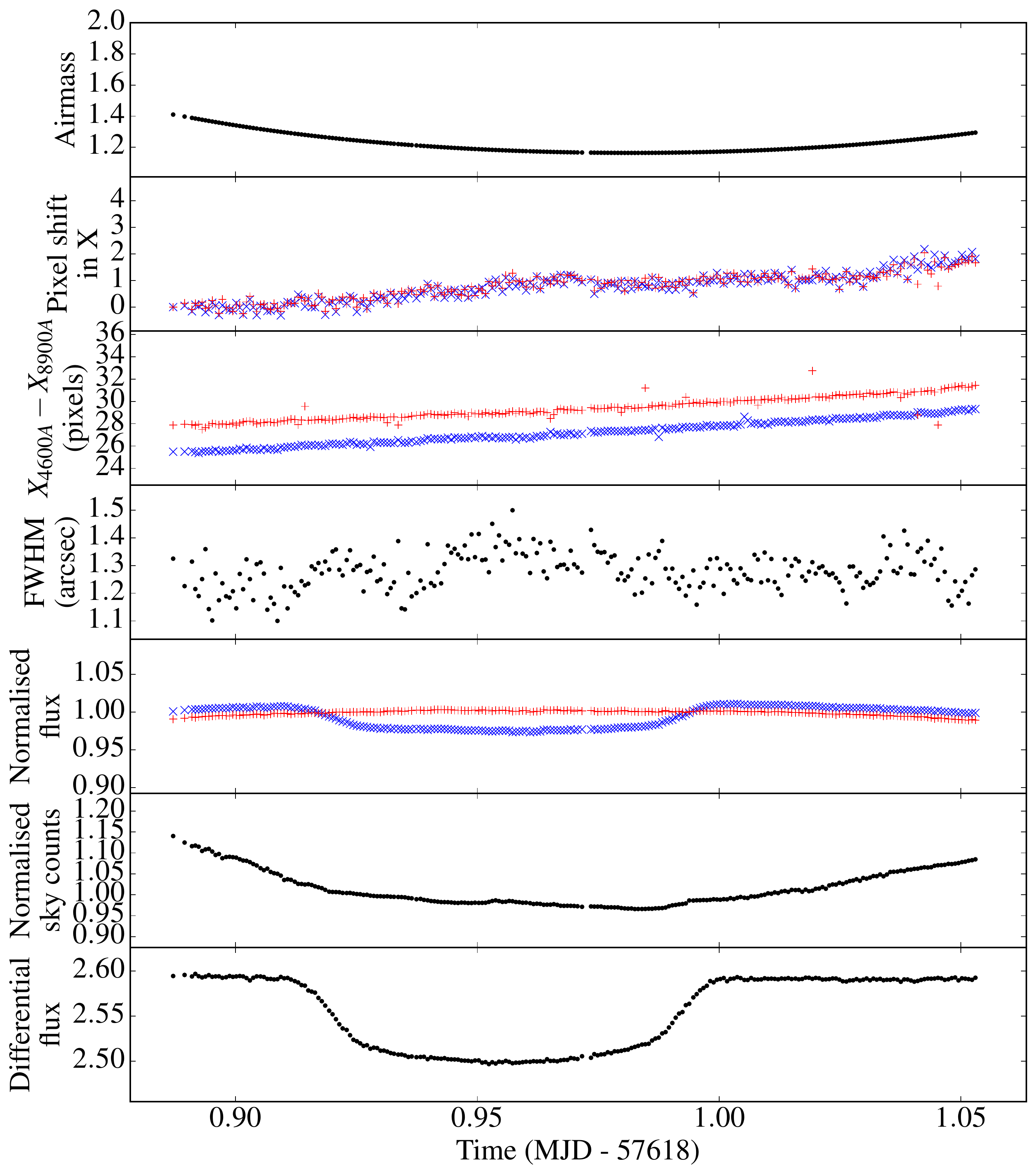}
\includegraphics[scale=0.3]{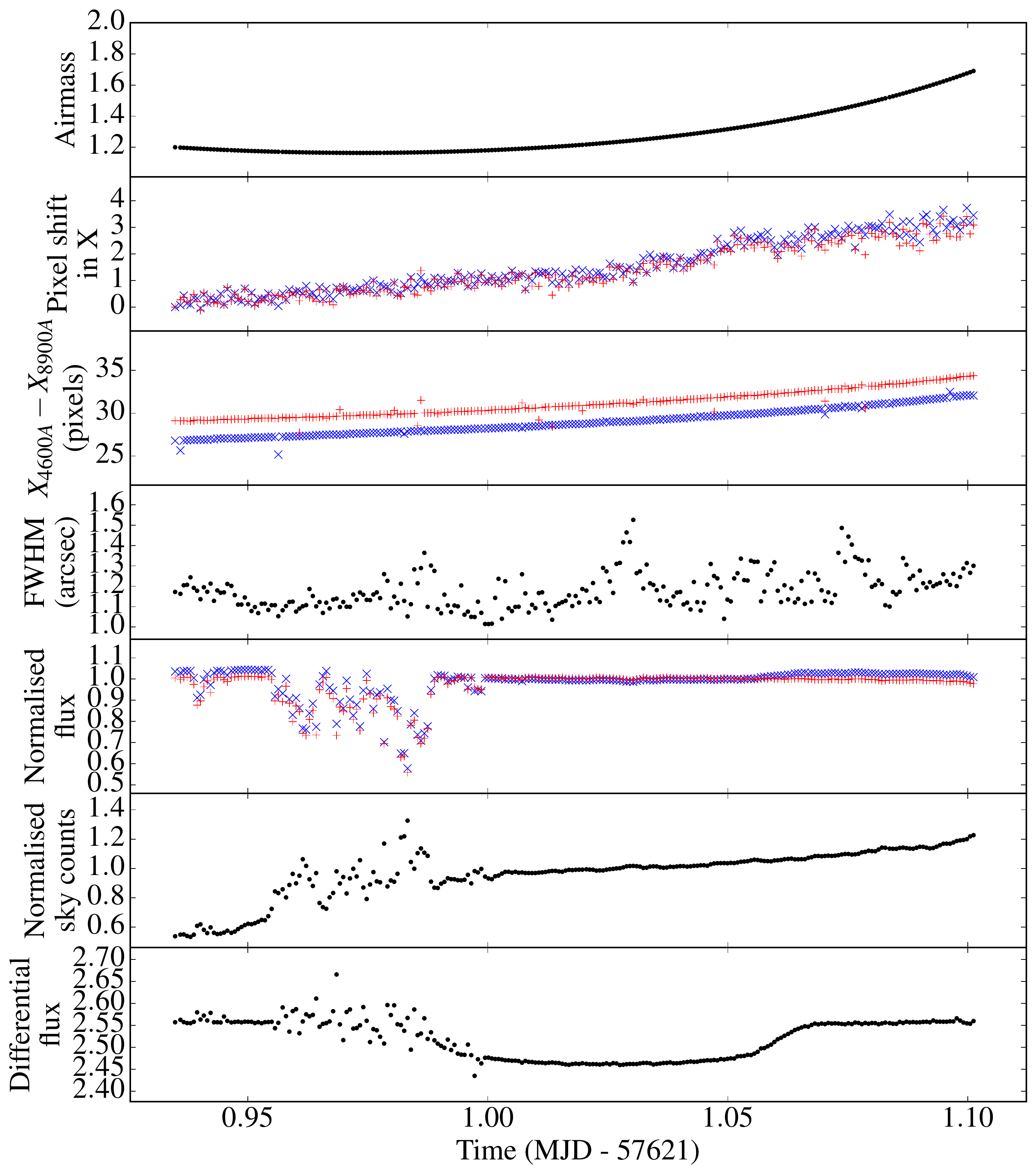}
\caption{Diagnostics of night 1 (left) and night 2 (right). Top panels: the variation of airmass across the nights. Second panels: the pixel shift in x of the target (blue crosses) and comparison (red pluses) traces. Third panels: Rotation of the target (blue crosses) and comparison (red pluses) traces shown as the difference in x position at opposite ends of the traces. Fourth panels: the variation of the full width at half maximum of the traces across the night. Fifth panels: the normalised light curves for the target (blue crosses) and comparison (red pluses). Sixth panels: the sky background across the night. Bottom panels: the differential white light curves.}
\label{fig:ancillary_plots}
\end{figure*}

Diagnostics of the extraction of the data for night 1 and night 2 are shown in Fig. \ref{fig:ancillary_plots}. The clouds in night 2 are clearly seen as drops in transmission in the raw light curves of the target and comparison (Fig. \ref{fig:ancillary_plots}, right-hand column, fifth panel). Using the raw light curves, we removed frames corresponding to drops in the transmission before further analysis. Fig. \ref{fig:wl_fits} shows the white light curve for night 2 once the cloud-affected frames had been removed, leaving 174 frames.

With spectra extracted for each frame, we aligned the spectra in pixel space and wavelength calibrated them following the procedure of \cite{Kirk2017}. The pixel alignment was done by cross-correlating strong features in each spectral frame with a reference frame individually for the target and comparison. For each frame this resulted in pixel shift as a function of position on the CCD for both the target and comparison, which we fitted with a third-order polynomial. Using this polynomial, we then resampled the target and comparison spectra onto the grid of the reference frame, using \textsc{pysynphot}\footnote{https://pysynphot.readthedocs.io/en/latest/} which conserves flux. To wavelength calibrate, we constructed an arc solution for both the target and comparison. Although the arc frames were taken with a different slit to the science images, and therefore could not be used for absolute wavelength calibration, they did provide a useful starting point. The final wavelength calibration was performed using stellar and telluric lines in the spectra of the target and comparison.

\begin{figure}
\centering
\includegraphics[scale=0.25]{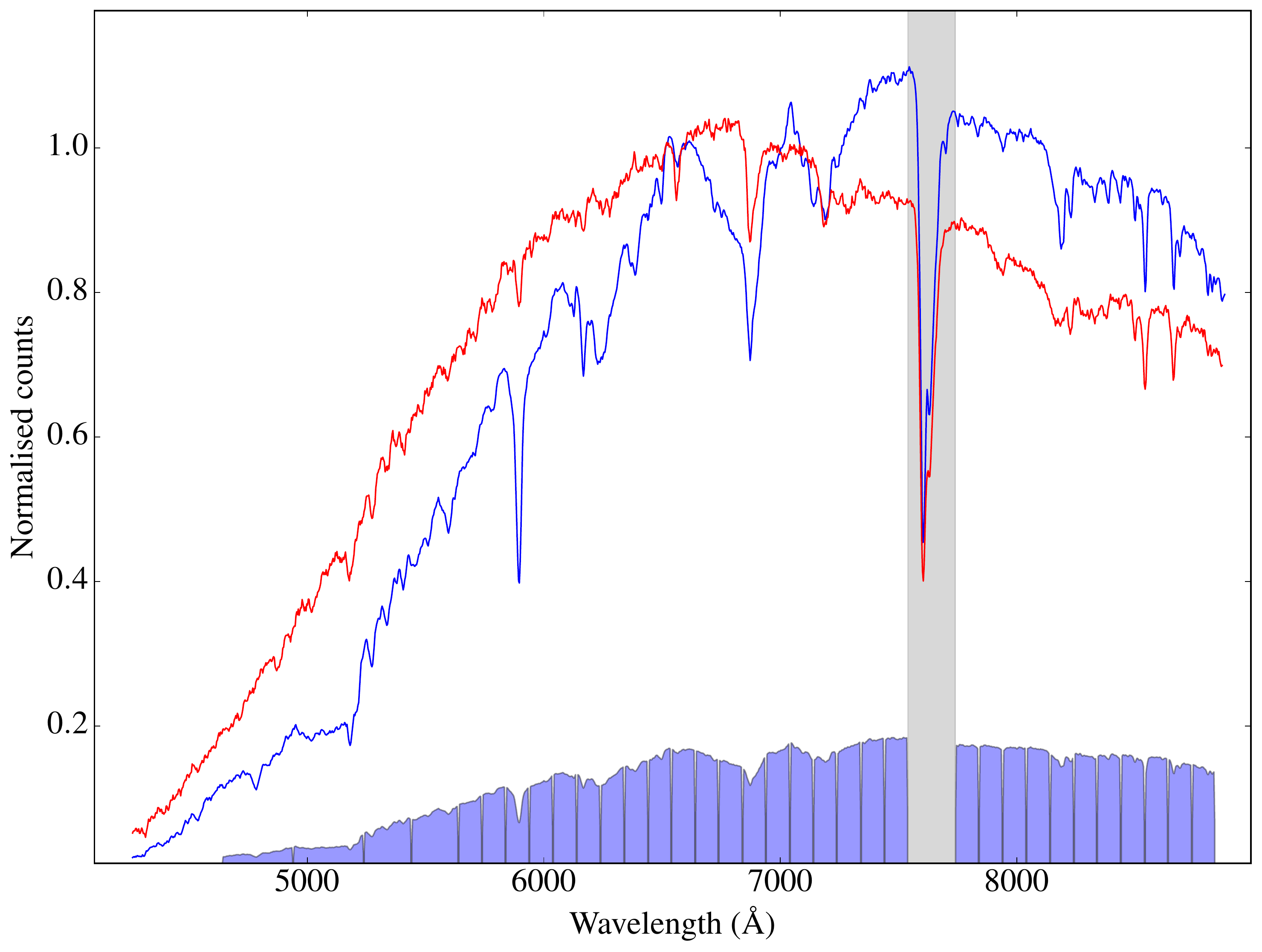}
\caption{Example extracted spectra of the target (blue) and comparison (red), normalised such that $F$(7000\,\AA) = 1.0. The lower filled boxes indicate the passbands used to create our wavelength-binned light curves. The grey region was excluded from our analysis.}
\label{fig:bin_locations}
\end{figure}

Following the wavelength calibration, we divided the spectra into 35 wavelength bins (Fig. \ref{fig:bin_locations}). The bluest four bins were made wider than the rest to increase the signal to noise ratio in these bins, given the red spectrum of the star and red sensitivity of ACAM. This led to two bins 300\,\AA ~wide, two bins 200\,\AA ~wide and 31 bins 100\,\AA ~wide. This width was chosen as a compromise between signal to noise and resolution. Similarly to \cite{Gibson2017}, we used a Tukey window to smooth the edges of the bins to avoid problems arising from sharp bin edges (Fig. \ref{fig:bin_locations}). We chose to ignore the region containing the strong telluric oxygen feature at $\sim7600$\,\AA ~due to increased noise in this binned light curve. While there were other telluric features present in the spectra, these were weaker and the light curves containing these features were not significantly worse than neighbouring bins.

\section{Data Analysis}

\subsection{Fitting the white light curves}
\label{sec:wl_fits}

Following extraction of the spectra, we created normalised, differential, white light curves for both nights by integrating the spectra for each frame. We then fitted the white light curves from both nights together with analytic transit light curves \citep{MandelAgol} with a quadratic limb darkening law. As with \cite{Kirk2017}, we chose to include a Gaussian process (GP) to model the red noise in the data and obtain robust errors for each of our fitted parameters. This was performed using the \textsc{george} \textsc{python} package \citep{george}. 

GPs have been used many times in the literature and have been shown to be powerful in modelling correlated noise in data (e.g. \citealt{Gibson2012}; \citealt{Evans2015}; \citealt{Kirk2017}; \citealt{Louden2017}). To model the red noise in our light curves, we chose to use a Mat\'{e}rn 3/2 kernel, defined by the length scale $\tau$ and the amplitude $a$. We also included a white noise kernel, defined by the variance $\sigma$.

The parameters defining the model fitted to both white light curves were the inclination $i$, the ratio of semi-major axis to stellar radius $a/R_*$, the time of mid-transit $T_c$, the planet to star radius ratio $R_P/R_*$, the quadratic limb darkening coefficients $u1$ and $u2$, and the three parameters defining the GP, $a$, $\tau$ and $\sigma$. $i$ and $a/R_*$ were shared between both nights whereas the remaining parameters were not. This resulted in a total of 16 free parameters in the white light curve fits.

A prior was placed on the limb-darkening coefficients such that $u1 + u2 \leq 1$, with no priors placed on the other transit parameters. Loose, uniform priors were placed on the GP hyperparameters to encourage convergence. 

To generate the starting values for the limb darkening coefficients, we made use of the limb darkening toolkit (\textsc{ldtk}; \citealt{LDTK}), which uses \textsc{phoenix} \citep{Husser2013} models to derive limb darkening coefficients and errors. For the stellar parameters, we used the values of \cite{Triaud2015}. For the starting values of the GP hyperparameters, we optimised these to the out of transit data, using a Nelder-Mead algorithm \citep{NelderMead}.

\begin{figure}
\centering
\includegraphics[scale=0.25]{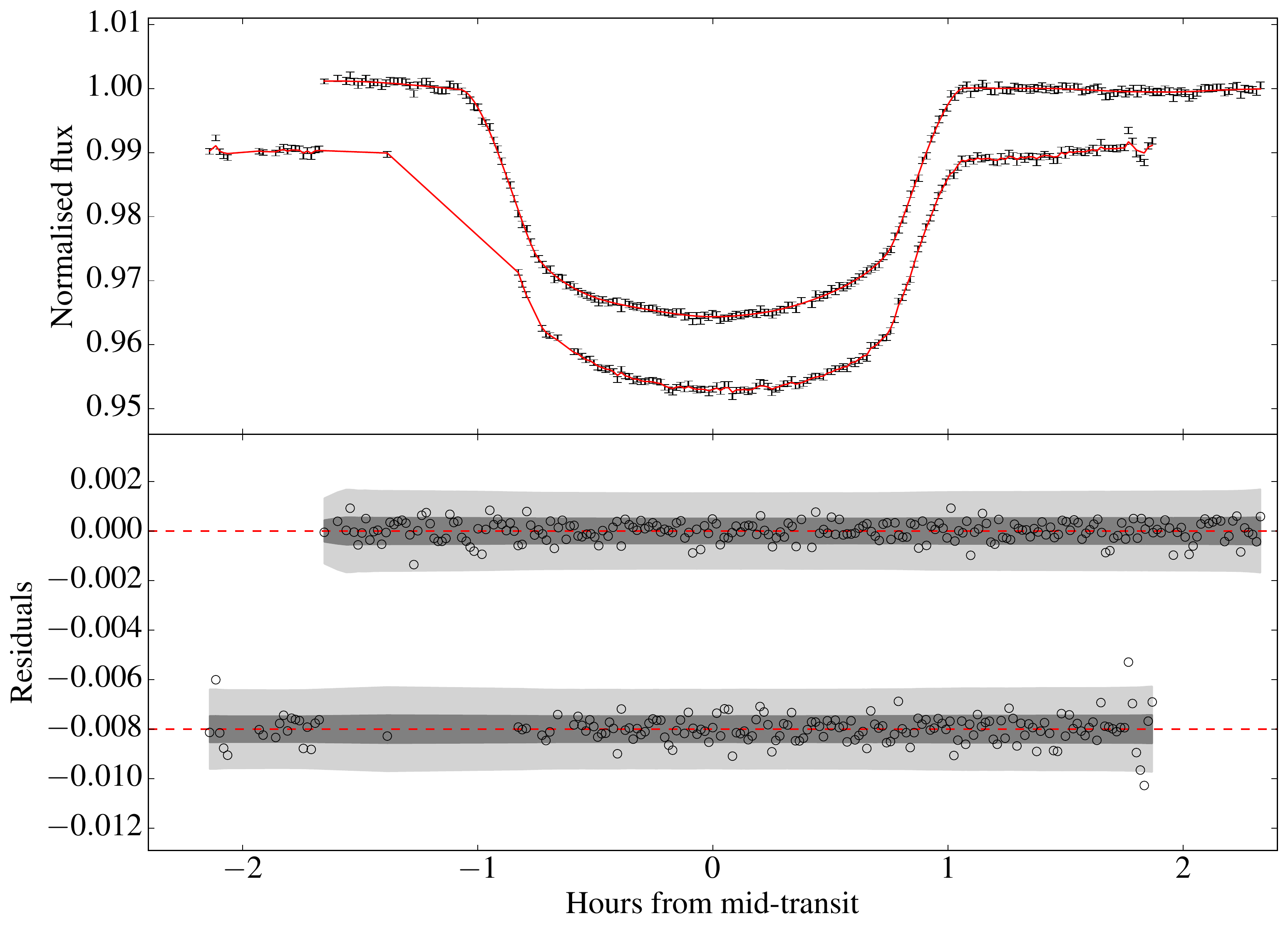}
\caption{Combined fit to the white light curves of nights 1 and 2 using a Gaussian process. Top panel: the fitted model to nights 1 and 2 is shown by the red line, with an offset in y of -0.01 applied for clarity. The gaps in night 2's light curve indicate the points that were clipped due to clouds. Bottom panel: The residuals to the fits in the top panel, offset by -0.008 for clarity. The dark grey and light grey regions indicate the 1 and 3$\sigma$ confidence intervals of the GP, respectively.}
\label{fig:wl_fits}
\end{figure}

With these starting values, we ran a Markov chain Monte Carlo (MCMC) to the white light curves of both nights together, using the \textsc{emcee} \textsc{python} package \citep{emcee}. An initial run was performed with 2000 steps and 160 walkers. Following this the walkers were resampled around the result from the first run and run again for 2000 steps with 160 walkers, with the first 1000 steps discarded as burn in. The resulting fit is shown in Fig. \ref{fig:wl_fits} with the parameter values in Table \ref{tab:wl_fit_results}.

\begin{table*}
\centering
\caption{Results from the fitting of the white light curves of nights 1 and 2 using a Gaussian process. Those parameters without an entry in the `Night 2' column were shared between the nights. We also include the results from \protect\cite{Sedaghati2017} and the discovery paper \protect\citep{Triaud2013}.}
\label{tab:wl_fit_results}
\begin{tabular}{|l|c|c|c|c|c|} \hline
Parameter & Night 1 & Night 2 & \protect\citealt{Sedaghati2017} & \protect\citealt{Triaud2013} \\ \hline

$a/R_*$ & $12.66^{+0.12}_{-0.11}$ & - & $12.0647 \pm 0.0099$ & $12.99 \pm 0.03$ \\

$i$ ($^\circ$) & $89.10^{+0.31}_{-0.22}$ & - & $88.90 \pm 0.06$  & $88.92^{+0.07}_{-0.12}$\\

$T_c$ (BJD) & $2457619.461954^{+0.000096}_{-0.000102}$  & $2457622.529914^{+0.000135}_{-0.000131}$ & $2456459.809578 \pm 0.000073$ & $2456125.417512^{+0.000067}_{-0.000052}$\\

$R_P/R_*$ & $0.16975^{+0.00179}_{-0.00178}$ & $0.17302^{+0.00198}_{-0.00235}$ & $0.17386 \pm 0.00030$ & $0.17126^{+0.00017}_{-0.00026}$\\

$u1$ & $0.47^{+0.09}_{-0.10}$ & $0.48 \pm 0.09$ & $0.491 \pm 0.238^a$ & - \\

$u2$ & $0.23 \pm 0.17 $ & $0.21^{+0.18}_{-0.17}$ & $0.485 \pm 0.198^a$ & - \\ \hline

\multicolumn{5}{l}{$^a$Note that these limb darkening coefficients are calculated over a different wavelength range to the study presented here.}

\end{tabular}
\end{table*}

\subsection{Wavelength-binned light curve fitting}

The fitting of the white light curves allowed us to derive a common set of system parameters and independent common noise models from each of the nights' light curves. With these we were able to fit wavelength-binned light curves for each night, after removing the common mode systematics model from each light curve and holding the system parameters ($a/R_*$ and $i$) and the time of mid transit $T_C$ fixed to the results from the white light curves. 

We again used a GP defined by a Mat\'{e}rn 3/2 kernel and a white noise kernel in order to account for the noise in the data. The remaining parameters defining the wavelength dependent models were the same as for the white light curve, resulting in 6 free parameters per light curve ($R_P/R_*$, $u1$, $u2$, $a$, $\tau$ and $\sigma$). For the limb darkening parameters, we used the log-likelihood evaluation of \textsc{ldtk} \citep{LDTK} in our fitting, and recovered values that were consistent with those that we measured from the white light curve.

We began by fitting an analytic transit light curve model with a cubic-in-time polynomial to remove any points that deviated by greater than $4\sigma$ from the fit. This typically clipped at most 1 to 2 points per light curve. As with the white light curves, we then optimised the GP hyperparameters to the out of transit data using a Nelder-Mead algorithm. Following these steps, we marginalised over all 6 free parameters with an MCMC. This was run for each of the light curves, with 2000 steps and 60 walkers. A second run was then performed for each, with a small perturbation added to the values resulting from the first run and again for 2000 steps with 60 walkers. 

The fitted light curves are shown in Fig. \ref{fig:wb_fits_night1} and Fig. \ref{fig:wb_fits_night2} with the resulting $R_P/R_*$ values listed in Table \ref{tab:bin_fits} and transmission spectrum plotted in Fig. \ref{fig:trans_spec1}. In Table \ref{tab:bin_fits} we also include the error in the combined $R_P/R_*$ in terms of the atmospheric scale height $H$, which we calculate to be 206\,km, given an equilibrium temperature of 825\,K \citep{Triaud2015}, a planetary surface gravity of 14.34\,ms$^{-1}$ \citep{Mancini2014_w80} and assuming the mean molecular mass to be 2.3 times the mass of a proton. 

Our results have a mean error across the transmission spectrum of 2.5 scale heights (0.00125\,$R_P/R_*$) which represents a good precision given the relatively small scale height of WASP-80b.

\begin{table*}
\centering
\caption{Results of the wavelength bin fits shown in Figs. \ref{fig:wb_fits_night1} \& \ref{fig:wb_fits_night2}.}
\label{tab:bin_fits}

\begin{tabular}{|c|c|c|c|c|} \hline
Wavelength & Night 1 & Night 2 & Combined & Error  \\ 
bin & $R_P/R_*$ & $R_P/R_*$ & $R_P/R_*$  & in scale heights \\ \hline

4640 -- 4940\AA & $0.17509^{+0.00084}_{-0.00085}$ & $0.17119^{+0.00332}_{-0.00227}$ & $0.17474 \pm 0.00081$ & 1.6 \\
4940 -- 5240\AA & $0.17126^{+0.00439}_{-0.00475}$ & $0.17162^{+0.00228}_{-0.00195}$ & $0.17147 \pm 0.00192$ & 3.7 \\
5240 -- 5440\AA & $0.17354^{+0.00399}_{-0.00496}$ & $0.17324^{+0.00295}_{-0.00280}$ & $0.17339 \pm 0.00242$ & 4.7 \\
5440 -- 5640\AA & $0.17082^{+0.00694}_{-0.00716}$ & $0.17326^{+0.00289}_{-0.00325}$ & $0.17300 \pm 0.00282$ & 5.5 \\
5640 -- 5740\AA & $0.17594^{+0.00470}_{-0.00570}$ & $0.17267^{+0.00244}_{-0.00238}$ & $0.17329 \pm 0.00219$ & 4.3 \\
5740 -- 5840\AA & $0.17606^{+0.00245}_{-0.00219}$ & $0.17328^{+0.00208}_{-0.00197}$ & $0.17441 \pm 0.00153$ & 3.0 \\
5840 -- 5940\AA & $0.17062^{+0.00397}_{-0.00287}$ & $0.17144^{+0.00295}_{-0.00323}$ & $0.17093 \pm 0.00230$ & 4.5 \\
5940 -- 6040\AA & $0.17726^{+0.00320}_{-0.00346}$ & $0.17249^{+0.00279}_{-0.00278}$ & $0.17450 \pm 0.00213$ & 4.1 \\
6040 -- 6140\AA & $0.17050^{+0.00325}_{-0.00312}$ & $0.17315^{+0.00165}_{-0.00155}$ & $0.17257 \pm 0.00143$ & 2.8 \\
6140 -- 6240\AA & $0.16903^{+0.00340}_{-0.00334}$ & $0.17382^{+0.00133}_{-0.00122}$ & $0.17318 \pm 0.00119$ & 2.3 \\
6240 -- 6340\AA & $0.16675^{+0.00516}_{-0.00484}$ & $0.17074^{+0.00219}_{-0.00180}$ & $0.17004 \pm 0.00186$ & 3.6 \\
6340 -- 6440\AA & $0.16993^{+0.00199}_{-0.00244}$ & $0.17290^{+0.00173}_{-0.00180}$ & $0.17183 \pm 0.00138$ & 2.7 \\
6440 -- 6540\AA & $0.17216^{+0.00165}_{-0.00129}$ & $0.17167^{+0.00248}_{-0.00192}$ & $0.17184 \pm 0.00122$ & 2.4 \\
6540 -- 6640\AA & $0.17171^{+0.00108}_{-0.00091}$ & $0.17150^{+0.00208}_{-0.00280}$ & $0.17167 \pm 0.00092$ & 1.8 \\
6640 -- 6740\AA & $0.16922^{+0.00178}_{-0.00226}$ & $0.17267^{+0.00236}_{-0.00217}$ & $0.17082 \pm 0.00151$ & 2.9 \\
6740 -- 6840\AA & $0.17020^{+0.00154}_{-0.00156}$ & $0.17261^{+0.00185}_{-0.00178}$ & $0.17121 \pm 0.00118$ & 2.3 \\
6840 -- 6940\AA & $0.17209^{+0.00149}_{-0.00152}$ & $0.17206^{+0.00213}_{-0.00181}$ & $0.17204 \pm 0.00120$ & 2.3 \\
6940 -- 7040\AA & $0.17241^{+0.00106}_{-0.00115}$ & $0.17003^{+0.00163}_{-0.00197}$ & $0.17182 \pm 0.00094$ & 1.8 \\
7040 -- 7140\AA & $0.17110^{+0.00097}_{-0.00189}$ & $0.17059^{+0.00057}_{-0.00051}$ & $0.17067 \pm 0.00051$ & 1.0 \\
7140 -- 7240\AA & $0.17496^{+0.00165}_{-0.00146}$ & $0.17119^{+0.00244}_{-0.00281}$ & $0.17396 \pm 0.00134$ & 2.6 \\
7240 -- 7340\AA & $0.17356^{+0.00143}_{-0.00155}$ & $0.17016^{+0.00151}_{-0.00134}$ & $0.17177 \pm 0.00103$ & 2.0 \\
7340 -- 7440\AA & $0.17186^{+0.00136}_{-0.00147}$ & $0.16880^{+0.00246}_{-0.00122}$ & $0.17060 \pm 0.00113$ & 2.2 \\
7440 -- 7540\AA & $0.17210^{+0.00168}_{-0.00114}$ & $0.16949^{+0.00150}_{-0.00127}$ & $0.17061 \pm 0.00099$ & 1.9 \\
7740 -- 7840\AA & $0.17179^{+0.00153}_{-0.00163}$ & $0.17007^{+0.00056}_{-0.00051}$ & $0.17023 \pm 0.00051$ & 1.0  \\
7840 -- 7940\AA & $0.17365^{+0.00133}_{-0.00150}$ & $0.16950^{+0.00067}_{-0.00060}$ & $0.17018 \pm 0.00058$ & 1.1  \\
7940 -- 8040\AA & $0.17140^{+0.00146}_{-0.00140}$ & $0.16984^{+0.00060}_{-0.00054}$ & $0.17003 \pm 0.00053$ & 1.0  \\
8040 -- 8140\AA & $0.17171^{+0.00149}_{-0.00155}$ & $0.17107^{+0.00063}_{-0.00058}$ & $0.17114 \pm 0.00056$ & 1.1  \\
8140 -- 8240\AA & $0.17087^{+0.00208}_{-0.00234}$ & $0.16923^{+0.00101}_{-0.00091}$ & $0.16947 \pm 0.00088$ & 1.7  \\
8240 -- 8340\AA & $0.17223^{+0.00152}_{-0.00160}$ & $0.17070^{+0.00093}_{-0.00082}$ & $0.17104 \pm 0.00076$ & 1.5  \\
8340 -- 8440\AA & $0.17239^{+0.00179}_{-0.00156}$ & $0.17058^{+0.00088}_{-0.00088}$ & $0.17095 \pm 0.00078$ & 1.5  \\
8440 -- 8540\AA & $0.17130^{+0.00152}_{-0.00161}$ & $0.17460^{+0.00113}_{-0.00091}$ & $0.17356 \pm 0.00086$ & 1.7  \\
8540 -- 8640\AA & $0.17451^{+0.00188}_{-0.00196}$ & $0.17353^{+0.00175}_{-0.00152}$ & $0.17390 \pm 0.00124$ & 2.4  \\
8640 -- 8740\AA & $0.17040^{+0.00154}_{-0.00156}$ & $0.17326^{+0.00127}_{-0.00155}$ & $0.17202 \pm 0.00104$ & 2.0  \\
8740 -- 8840\AA & $0.17094^{+0.00123}_{-0.00134}$ & $0.17488^{+0.00129}_{-0.00145}$ & $0.17284 \pm 0.00094$ & 1.8  \\

\hline
\end{tabular}
\end{table*}

\subsection{Noise sources in the wavelength binned light curves}

\begin{figure*}
\centering
\includegraphics[scale=0.25]{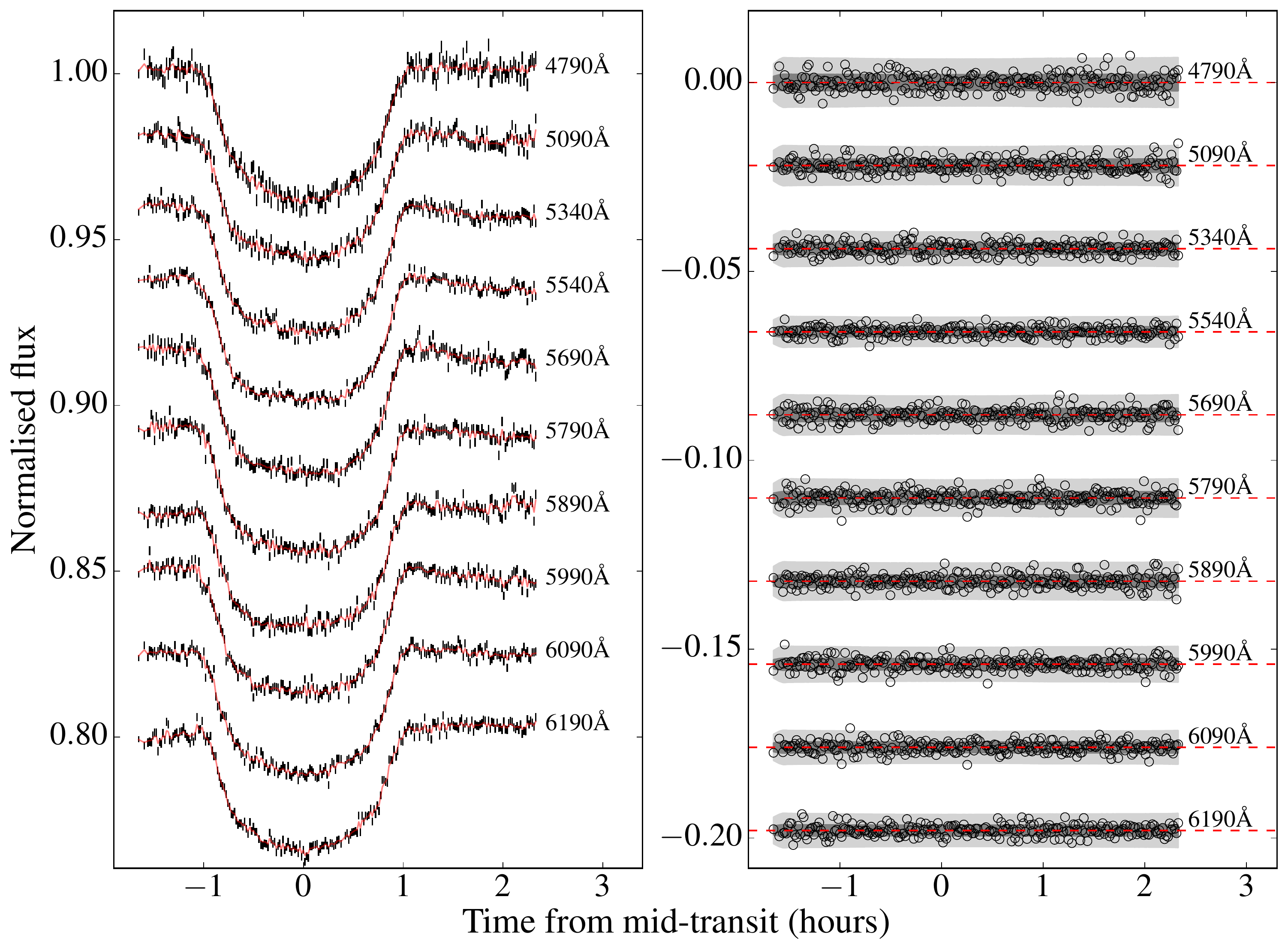}
\includegraphics[scale=0.25]{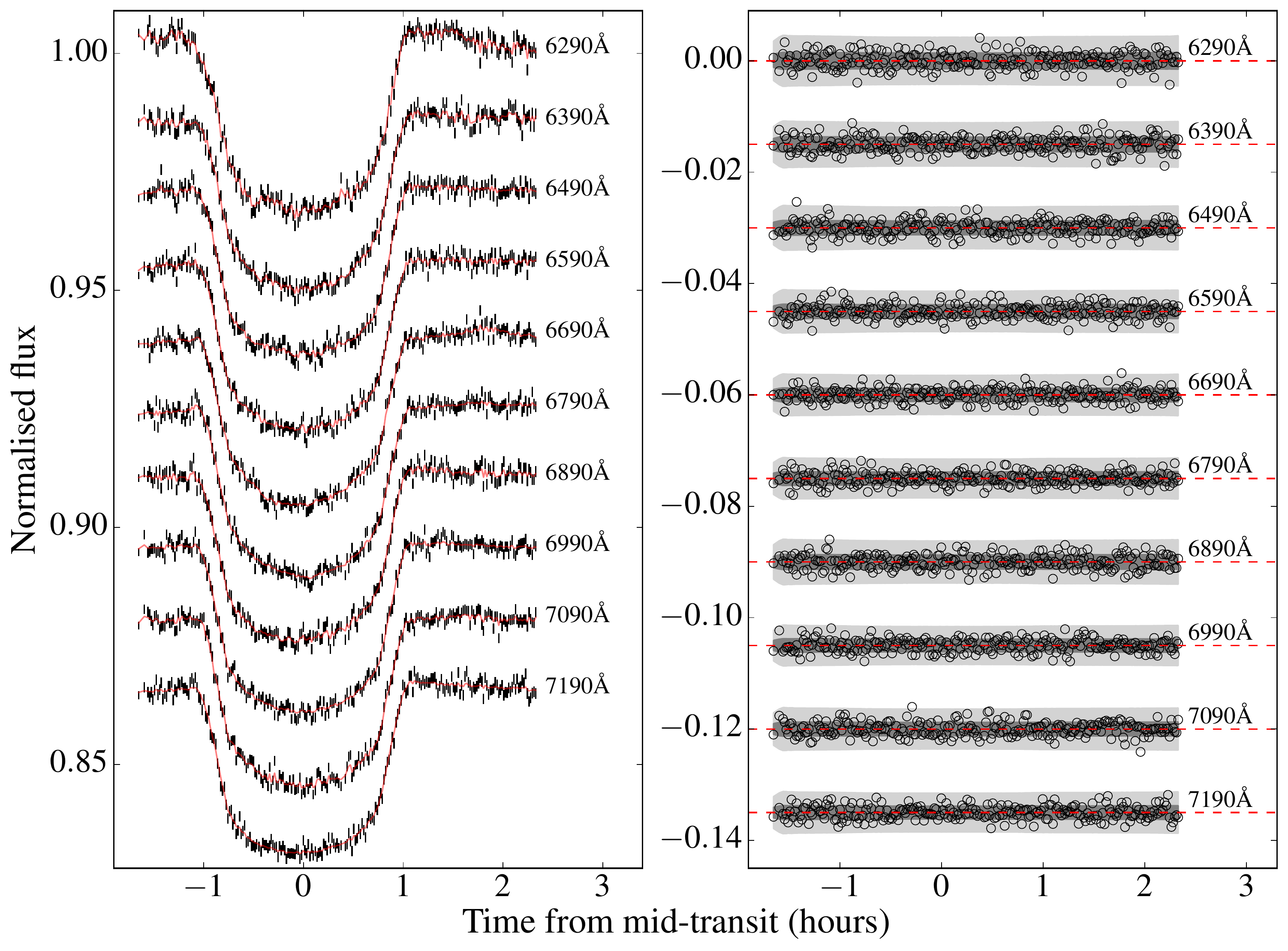}
\includegraphics[scale=0.25]{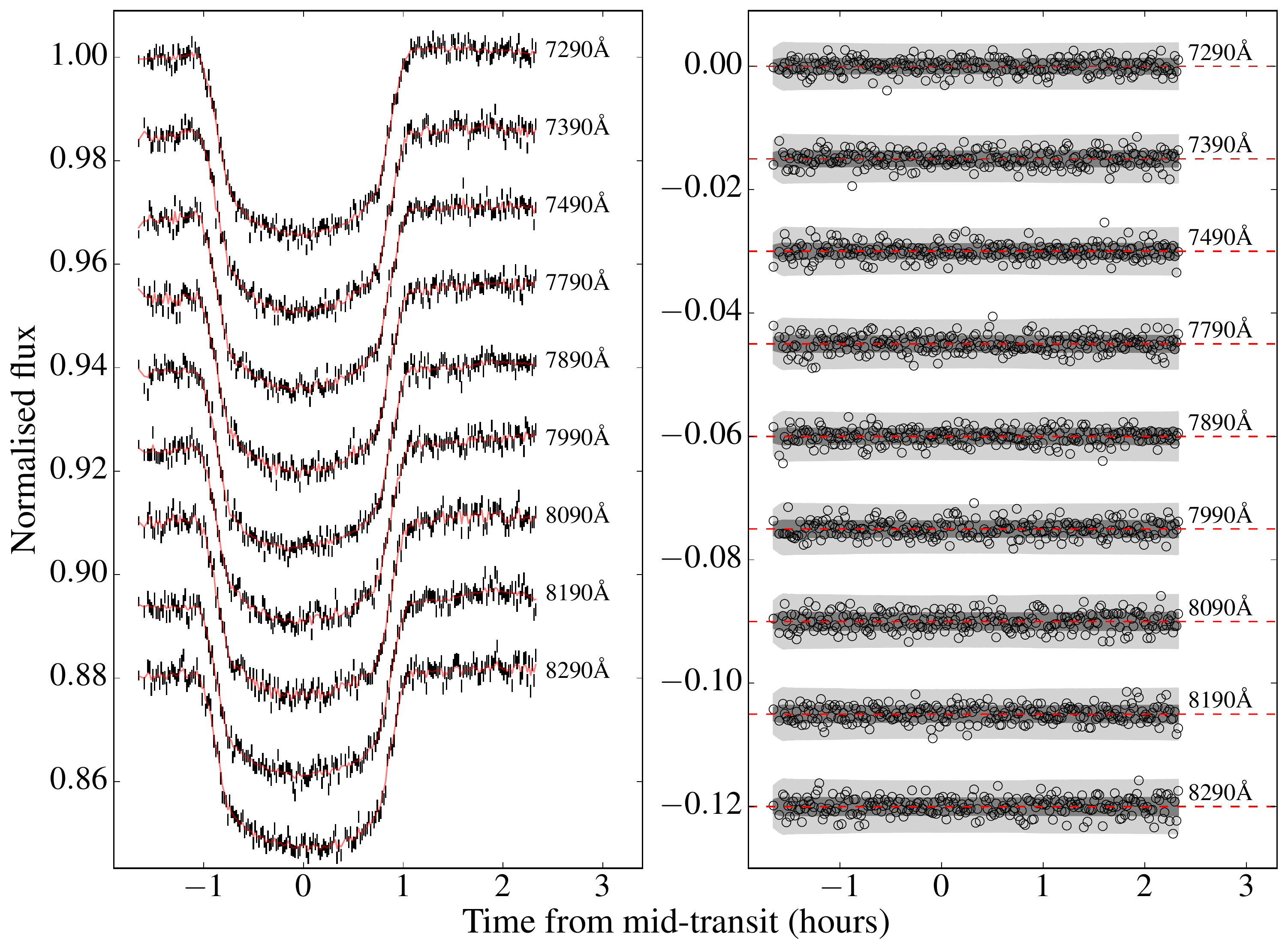}
\includegraphics[scale=0.25]{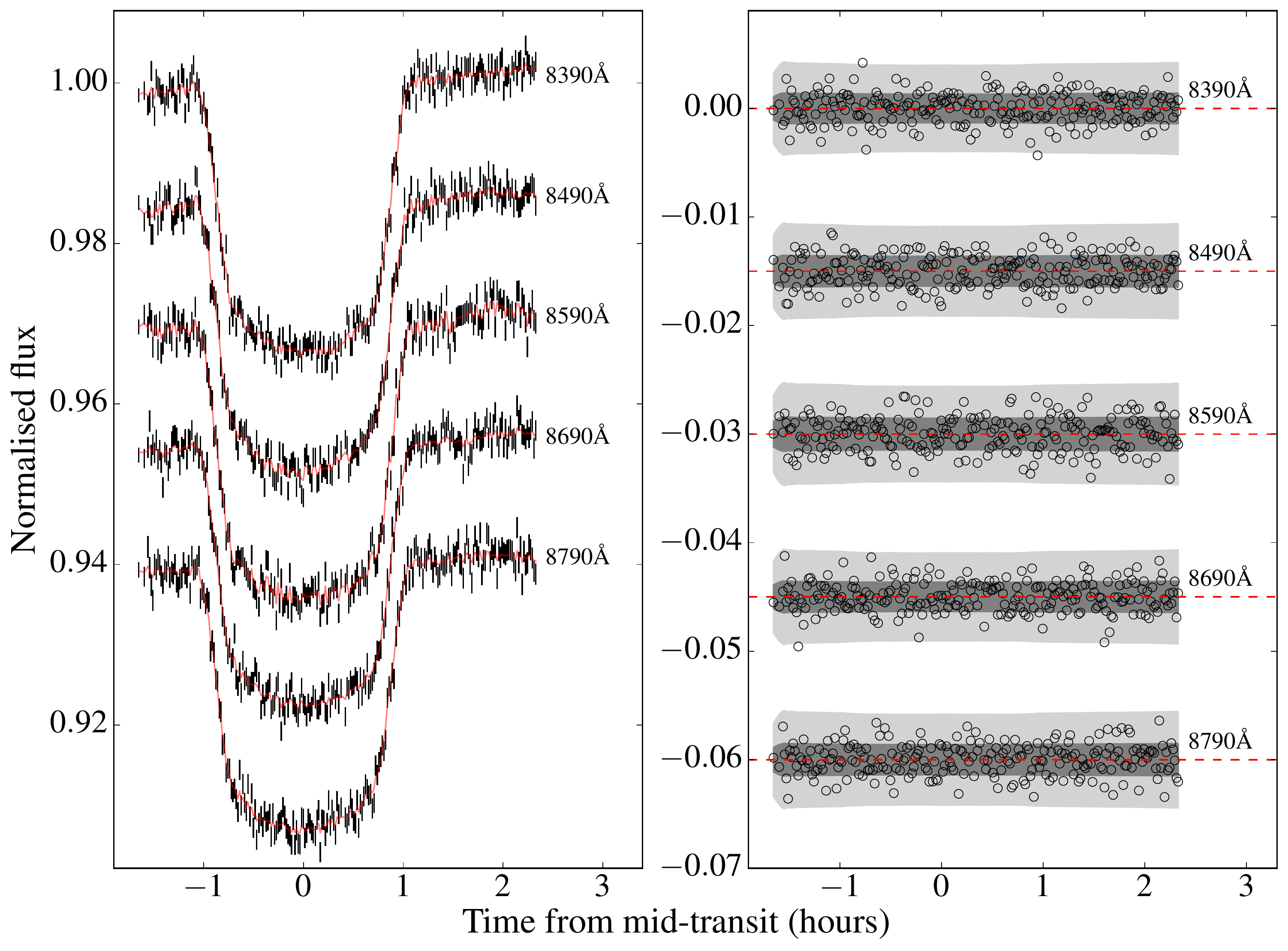}
\caption{Best fitting models to the wavelength binned light curves of night 1, which are offset for clarity. In each of the four plots, the left hand panel shows the data as black errorbars with the red line showing the best fitting model which included both an analytic transit light curve and a Gaussian process (GP). The right hand panels show the residuals to the best fitting model with the dark grey and light grey shaded regions indicating the 1 and 3$\sigma$ confidence intervals.}
\label{fig:wb_fits_night1}
\end{figure*}

\begin{figure*}
\centering
\includegraphics[scale=0.25]{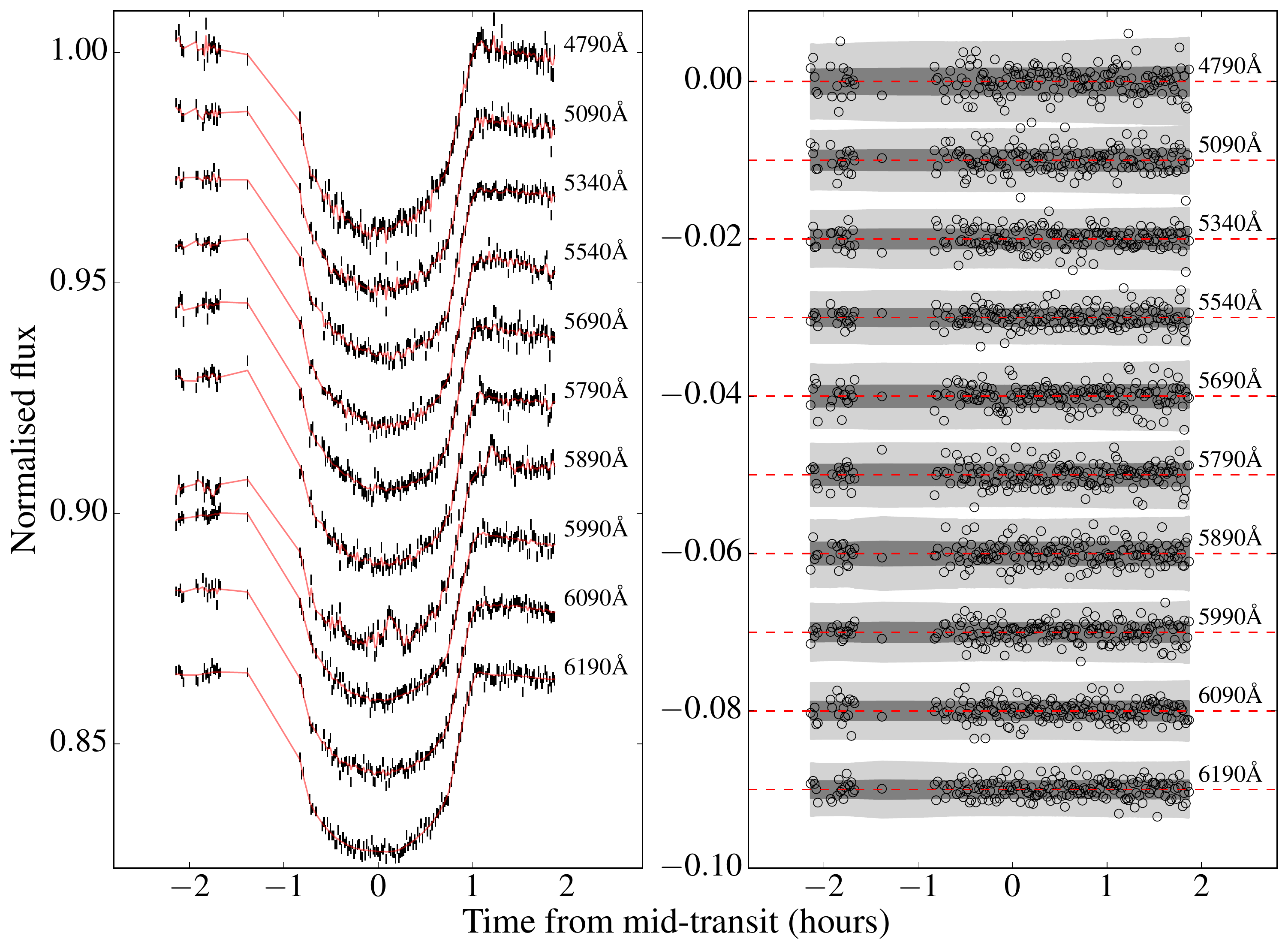}
\includegraphics[scale=0.25]{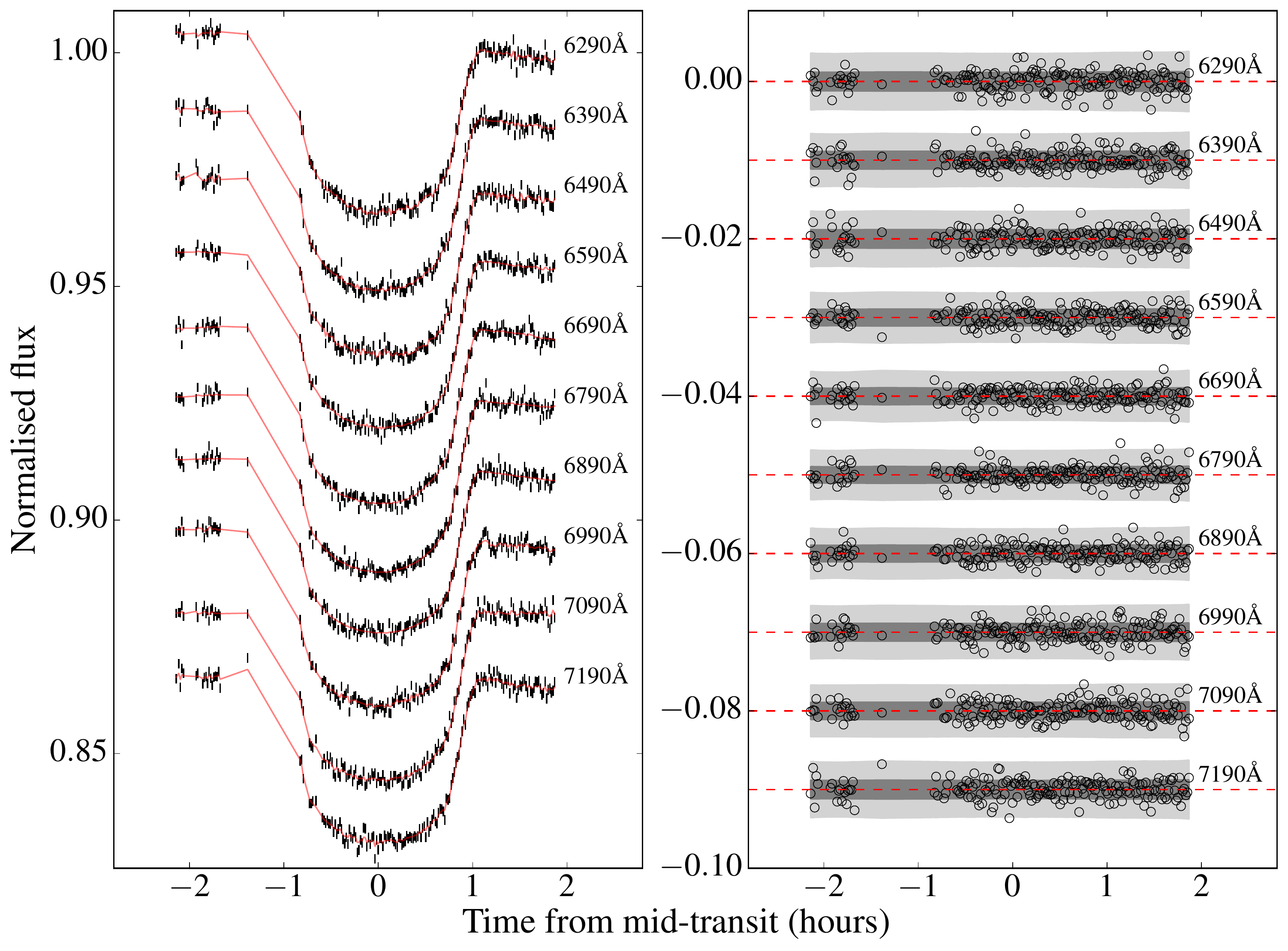}
\includegraphics[scale=0.25]{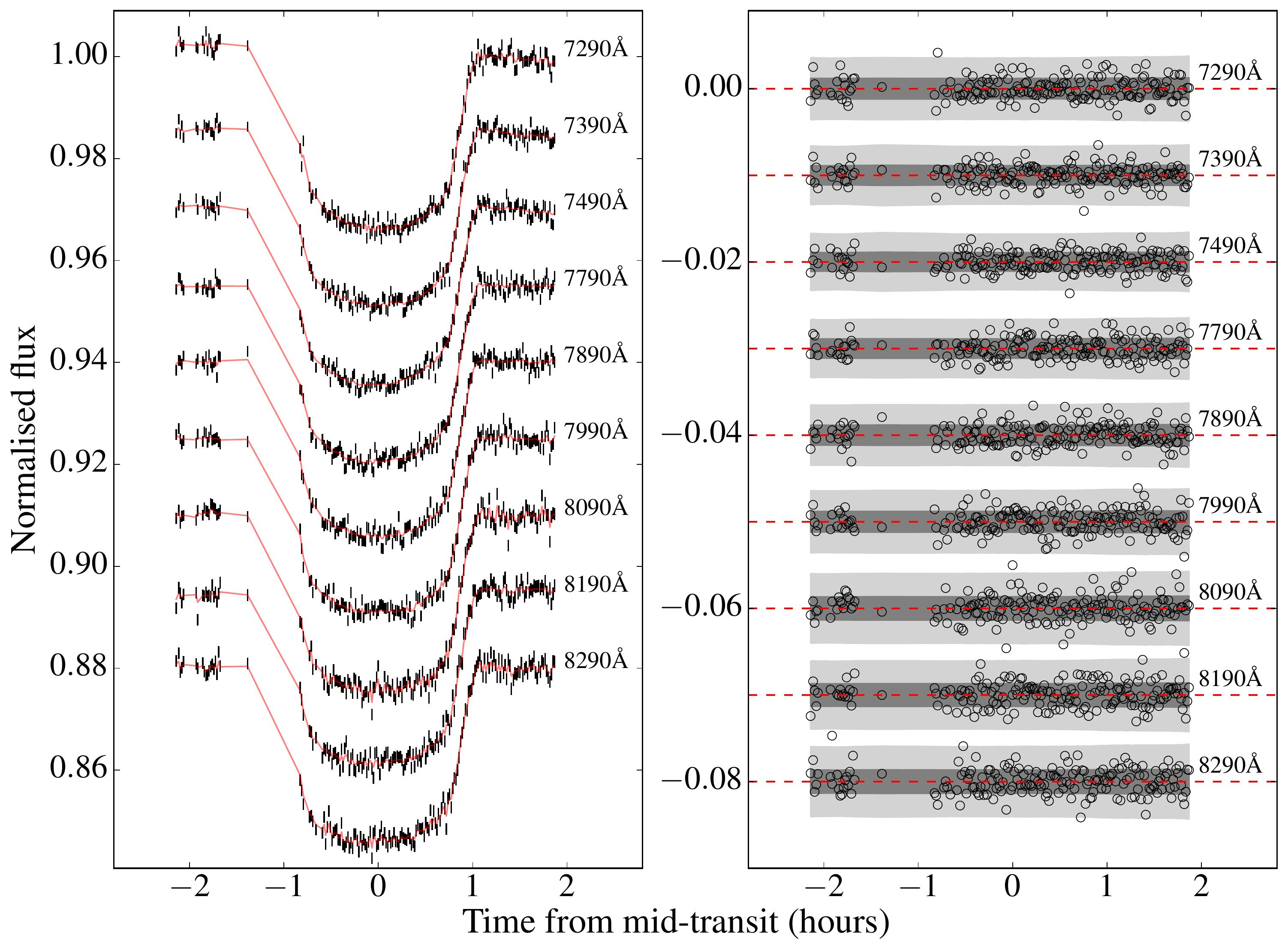}
\includegraphics[scale=0.25]{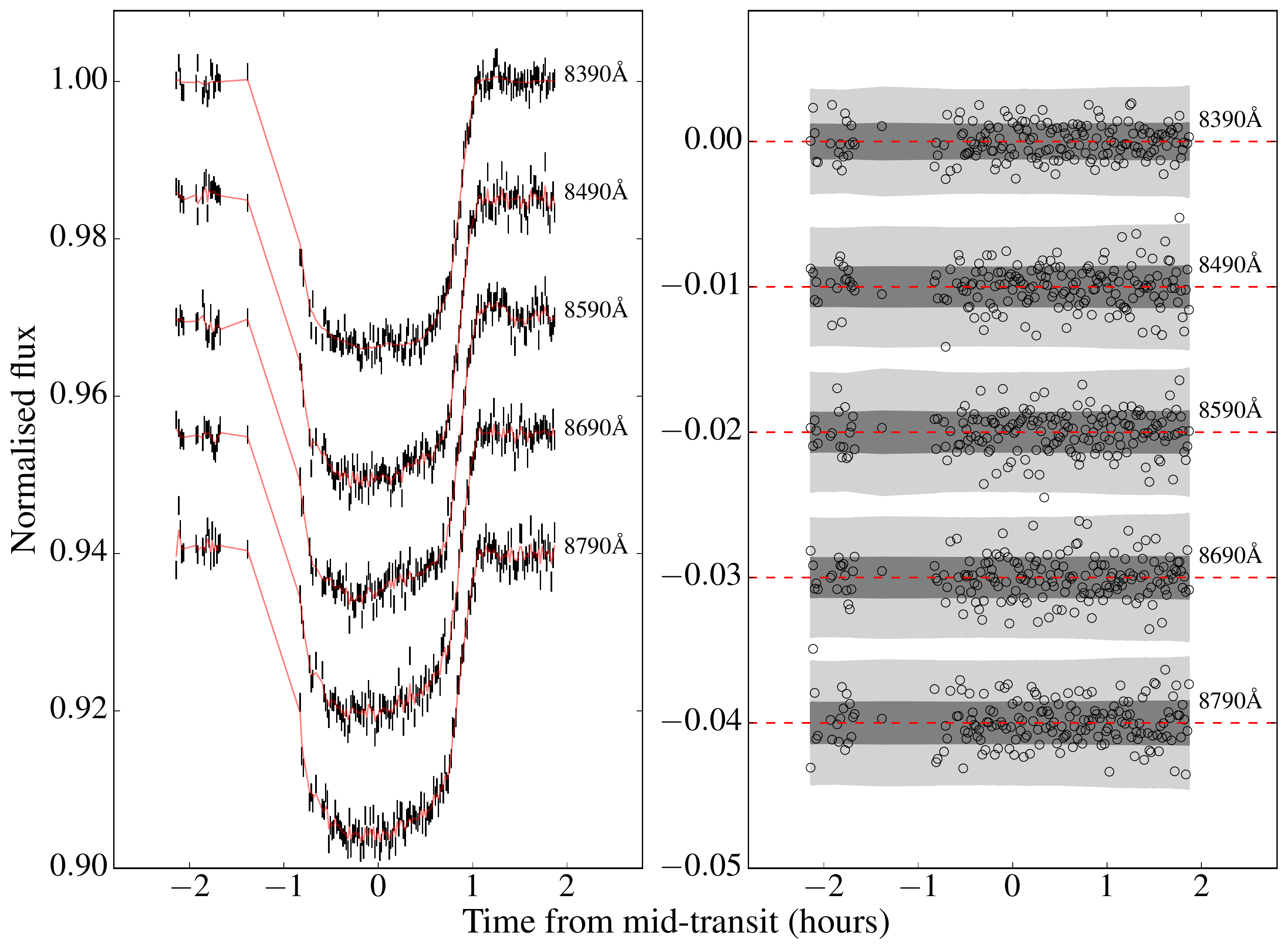}
\caption{The same plots as in Fig. \protect\ref{fig:wb_fits_night1} but for night 2's data.}
\label{fig:wb_fits_night2}
\end{figure*}

We looked for correlations between different diagnostics and noise features in our light curves. Aside from an airmass correlation, we did not see any correlations between trace position, FWHM or sky background with the wavelength-binned light curves whose systematics varied from bin to bin (Figs. \ref{fig:wb_fits_night1} and \ref{fig:wb_fits_night2}). 

We also considered that noise may have been introduced by our post-reduction processing. We tested our resampling method by constructing light curves from unresampled data and tested our wavelength solution by binning in pixels prior to assigning a wavelength, neither of which changed the dominant noise characteristics. 

We also experimented with a non-linearity correction from the ACAM instrument pages\footnote{http://www.ing.iac.es/engineering/detectors/auxcam\_fast.jpg} and re-ran our reduction pipeline but also found this had no significant effect. The effect of flat-fielding was also tested by running a reduction with a flat field using a tungsten lamp and a reduction without a flat field at all and we found that neither of these had a significant effect. 

\subsection{Transmission spectrum}
\label{sec:trans_spec_results}

\begin{figure*}
\centering
\includegraphics[scale=0.5]{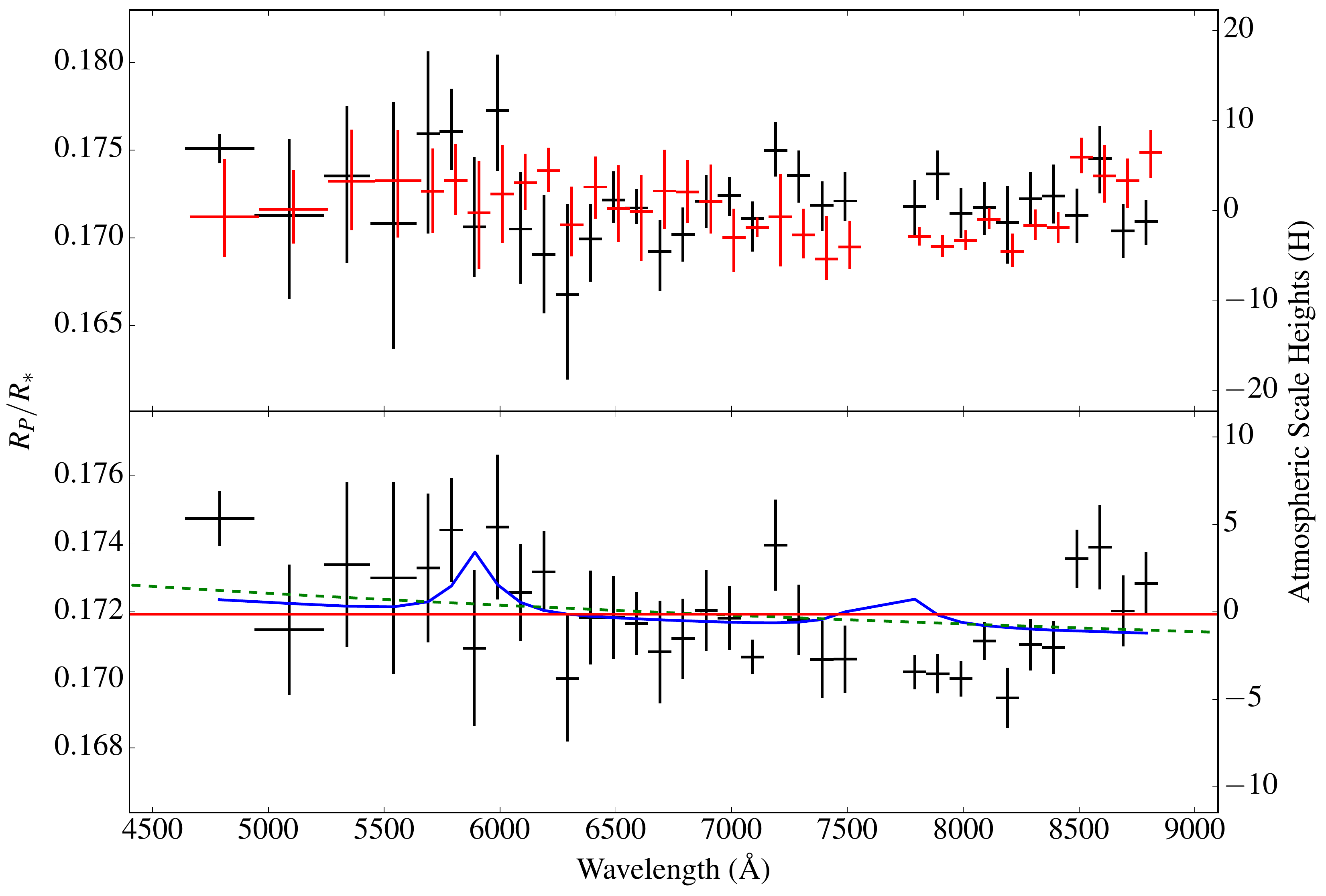}
\caption{Top panel: the transmission spectra resulting from the wavelength-binned fits to the data from night 1 (black error bars) and night 2 (red error bars). The data from night 2 are offset by +20\,\AA ~for clarity. Bottom panel: the transmission spectrum resulting from the weighted mean of the two individual nights (black error bars), along with a clear, $0.1 \times$ solar atmosphere (blue line) assuming an isothermal temperature pressure profile, a flat line indicating a cloudy atmosphere (red line), and a pure Rayleigh scattering model (green dashed line) at the equilibrium temperature of the planet (825\,K).}
\label{fig:trans_spec1}
\end{figure*}

Following the fitting of the wavelength binned light curves of both nights, we constructed individual transmission spectra for each night and a combined transmission spectrum from the weighted mean (Fig. \ref{fig:trans_spec1}). We have renormalised the transmission spectrum for the first night, given the difference in $R_P/R_*$ from the two nights. We note that the results for night 1 in Table \ref{tab:bin_fits} are with the offset applied.

In Fig. \ref{fig:trans_spec1}, we also plot our combined transmission spectrum along with a clear atmosphere model, a Rayleigh scattering slope, and a flat line indicating a grey opacity source such as clouds. The clear atmosphere model was generated using \textsc{exo-transmit} \citep{Kempton2017} and assuming an isothermal temperature pressure profile of 800\,K, with a metallicity of $0.1 \times$ solar.

\begin{table}
\centering
\caption{Goodness of fit of different model atmospheres to both our whole transmission spectrum (4640 -- 8840\AA, with 33 degrees of freedom) and the region overlapping with the study of \protect\cite{Sedaghati2017} (7440 -- 8840\AA, with 11 degrees of freedom). The 95\% confidence limit for a $\chi^2$ distribution with 33 degrees of freedom is 47.4 and for 11 degrees of freedom is 19.7. The models preferred in each region are shown in boldface.}
\label{tab:atmos_stats}
\begin{tabular}{|l|c|c|c|}\hline
& 4640 -- 8840\AA & 7440 -- 8840\AA \\
& $ \chi^2$ ($\chi^2_{\nu}$) & $\chi^2$ ($\chi^2_{\nu}$) \\ \hline
Flat line & 74.9 (2.27) & \textbf{32.1 (2.92)} \\
Rayleigh (T$_{\textrm{eq}}$) & \textbf{68.5 (2.08)} & 39.5 (3.59) \\
$0.1 \times$ solar, clear & 81.6 (2.47) & 50.6 (4.60) \\
$1 \times$ solar, clear & 71.6 (2.17) & 40.4 (3.67) \\
\hline
\end{tabular}
\end{table}

Table \ref{tab:atmos_stats} shows the goodness of fit of the model atmospheres considered. We find that while none of these models provide a satisfactory fit to the entire data set, a Rayleigh scattering slope at the equilibrium temperature of the planet (825\,K) is preferred.

\begin{figure}
\centering
\includegraphics[scale=0.25]{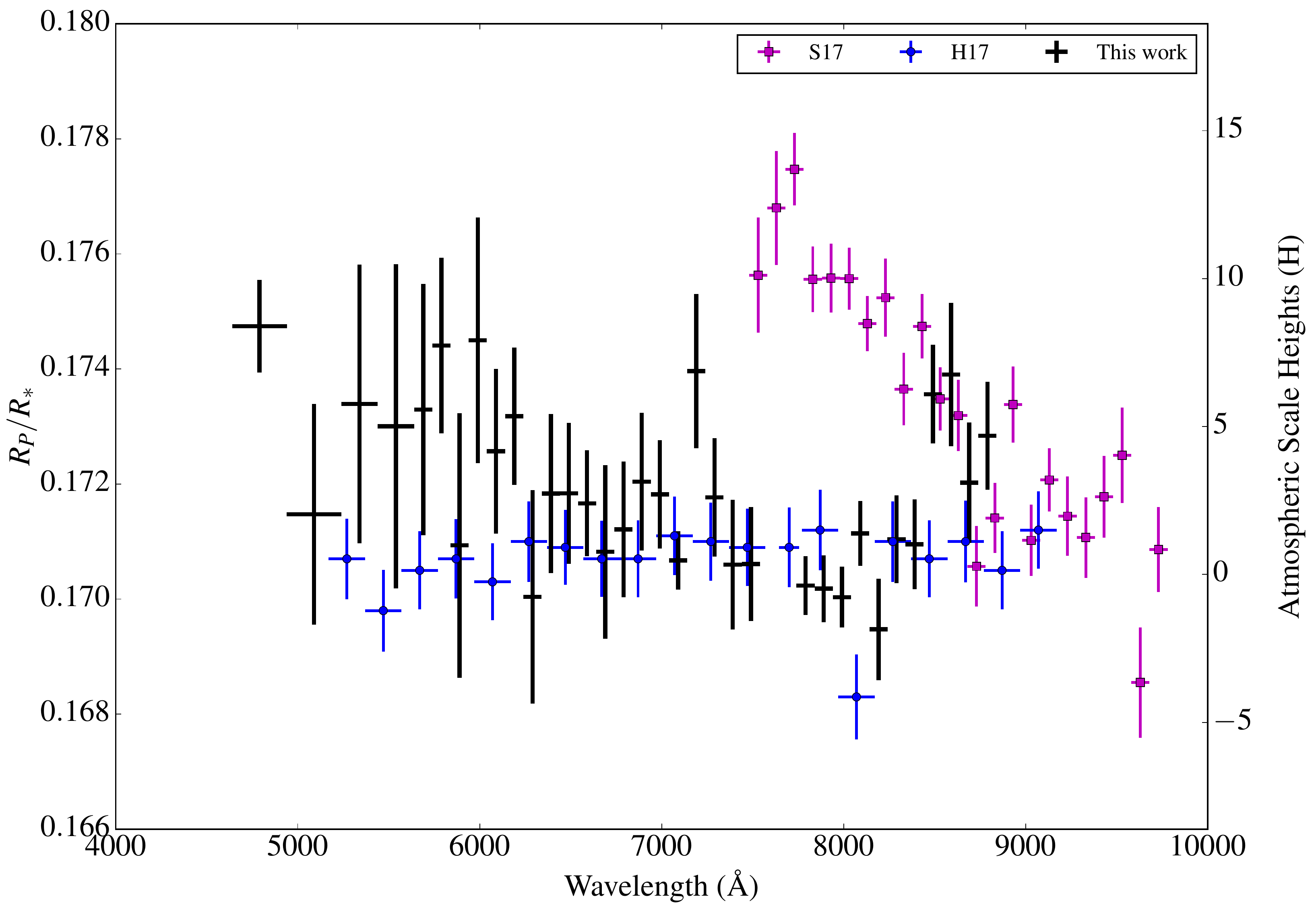}
\caption{Our two night combined transmission spectrum (black error bars) along with the transmission spectra of \protect\cite{Sedaghati2017} (magenta squares) and \protect\cite{Parviainen2017} (blue circles).}
\label{fig:trans_spec2}
\end{figure}

In Fig. \ref{fig:trans_spec2} we plot our combined transmission spectrum against the spectra of both \cite{Sedaghati2017} and \cite{Parviainen2017}. Our results clearly favour a transmission spectrum without broad potassium absorption and show no evidence of sodium at a resolution of 100\,\AA. When considering just the region that overlaps with the study of \cite{Sedaghati2017}, our transmission spectrum strongly favours a flat line over all the other models when comparing the Bayesian Information Criterion (which is a comparison of $\chi^2$ for the same degrees of freedom) for this region. However, the model atmospheres considered do not produce satisfactory fits in this region (Table \ref{tab:atmos_stats}). 

Fig. \ref{fig:trans_spec2} also provides a favourable comparison between the precision of our results using a 4-metre telescope with those from 8- and 10-metre telescopes (\citealt{Sedaghati2017}; \citealt{Parviainen2017}). It is clear that the precision of ground based studies is limited by systematics and not photon noise.

\section{Discussion}
\label{sec:discussion}

\subsection{Transmission spectrum}
\label{sec:trans_spec_discussion}

The transmission spectrum presented here is best represented by a Rayleigh scattering slope with no broad alkali metal absorption, suggesting the presence of a high altitude haze (Fig. \ref{fig:trans_spec1}, Table \ref{tab:atmos_stats}). It is inconsistent with the previously claimed detection of broad potassium absorption by \cite{Sedaghati2017}. 

In the study of \cite{Sedaghati2017}, the authors interpreted their transmission spectrum as evidence for the pressure-broadened wings of the potassium feature. This would suggest a clear atmosphere such as in WASP-39b (\citealt{Fischer2016}; \citealt{Sing2016}; \citealt{Nikolov2016}), as clouds and hazes have the effect of masking the wings of the alkali features (e.g. \citealt{Nikolov2015}; \citealt{Mallonn2015}; \citealt{Kirk2017}; \citealt{Louden2017}).  In this case, we might expect to see the broad wings of the sodium feature but we see no evidence for these either, nor do we detect the core at a resolution of 100\,\AA~ (Fig. \ref{fig:trans_spec1}). 

Alkali chlorides are expected to condense at around the temperature of WASP-80b (825\,K, \citealt{Mancini2014_w80}), which reduces the sodium and potassium in the upper atmosphere and gives rise to Rayleigh scattering \citep{Wakeford2015}. This scenario is consistent with our transmission spectrum as we do not detect broad sodium or potassium absorption, and find an atmosphere best represented by Rayleigh scattering (Fig. \ref{fig:trans_spec1}, Table \ref{tab:atmos_stats}). However, we might expect chlorides to produce a steeper slope than we observe \citep{Wakeford2015}.  

It is interesting to note that for smaller planets hazes appear more prominent for planets with temperatures $\lesssim900$\,K (\citealt{Morley2015}; \citealt{Crossfield2017}), which is the temperature range that both HAT-P-18b, for which we detected a Rayleigh scattering haze \citep{Kirk2017}, and WASP-80b fall within, although they probe a different parameter space and a different regime of atmospheric chemistry. Our results for these two relatively-cool hot Jupiters are also in line with \cite{Heng2016}'s tentative evidence that cooler planets are more likely to be cloudy.

Despite the presence of a haze, it is still possible to detect the narrow line core arising from higher altitudes in the planetary atmosphere such as has been done for WASP-52b \citep{Chen2017_w52}, HD~189733b  (\citealt{Pont2008}; \citealt{Sing2011}; \citealt{Huitson2012}; \citealt{Pont2013}; \citealt{Louden2015}) and HD~209458b (\citealt{Charbonneau2002}; \citealt{Sing2008a,Sing2008b}; \citealt{Snellen2008}; \citealt{Langland-Shula2009}; \citealt{Vidal-Madjar2011}; \citealt{Deming2013}), however these detections are often made in much narrower bins than the 100\,\AA ~bins used here. For this reason, we cannot rule out the presence of the narrow core of the sodium feature despite the absence of the broad wings. 

While we cannot be certain what the cause of the discrepancy is between our results and those of \cite{Sedaghati2017}, it is perhaps plausible that there were residual systematics in the VLT/FORS data of \cite{Sedaghati2017} associated with the Longitudinal Atmospheric Dispersion Corrector (LADC; \citealt{Avila1997}) despite the authors' thorough attempts to account for these. The observations of \cite{Sedaghati2017} were taken before the replacement of the anti-reflective coating of the LADC which was known to cause systematic errors \citep{Boffin2015}. We note that more recent observations by these authors, also using VLT/FORS, resulted in a ground-breaking detection of TiO in the atmosphere of WASP-19b but with the LADC left in park mode \citep{Sedaghati2017_wasp19}. We are also encouraged that ACAM is optically simpler than FORS with no atmospheric dispersion corrector.

In addition to possible systematics in the VLT/FORS data of \cite{Sedaghati2017}, we consider the effects of stellar activity in the next section.

\subsection{Stellar activity}

\begin{figure}
\centering
\includegraphics[scale=0.25]{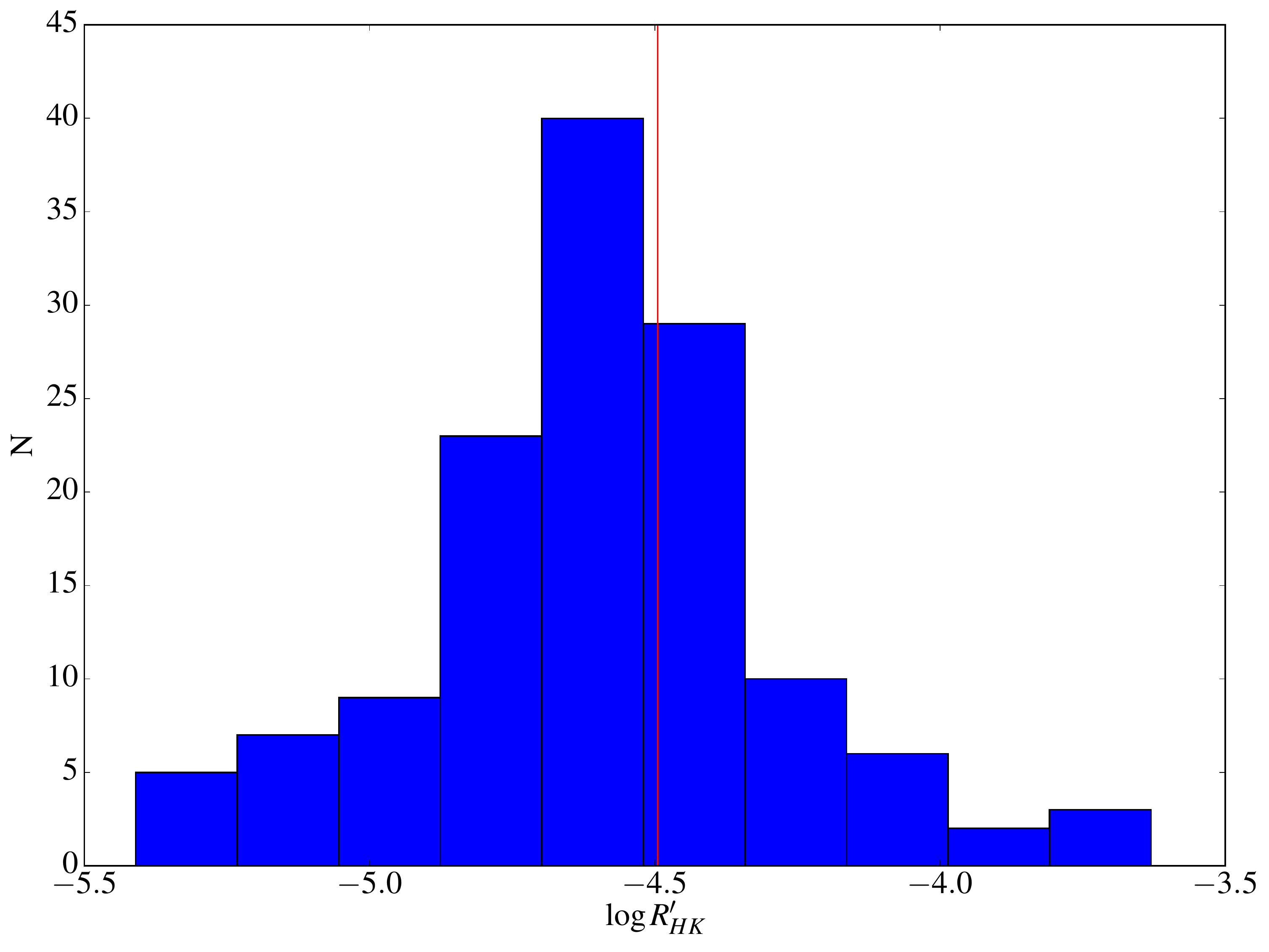}
\caption{A plot of $\log R'_{\textrm{HK}}$ values of K dwarfs displaying chromospheric features from the sample of \protect\cite{MartinezArnaiz2010}, who gathered chromospheric activity measurements of 371 F to K stars within 25\,pc. The red line indicates WASP-80's $\log R'_{\textrm{HK}}$.}
\label{fig:activity}
\end{figure}

Despite the lack of rotational modulation in its light curve, WASP-80 has a $\log R'_{\textrm{HK}}$ of -4.495, suggesting the star has a relatively high level of chromospheric activity \citep{Triaud2013,Mancini2014_w80}. However, WASP-80 has a spectral type between K7V and M0V \citep{Triaud2013} and when comparing this value to a sample of K stars with measured $\log R'_{\textrm{HK}}$ values, we see that WASP-80's $\log R'_{\textrm{HK}}$ is fairly typical (Fig. \ref{fig:activity}, values from \citealt{MartinezArnaiz2010}).

Given WASP-80's chromospheric activity, it is puzzling that no star spot crossings have been observed in any of the studies of WASP-80 to date. This would suggest that either WASP-80 is spot free or that there are a large number of spots continuously on its surface, at latitudes not occulted by the planet (as postulated by \citealt{Mancini2014_w80}).

The orbit of WASP-80b is aligned with the stellar spin axis \citep{Triaud2015} and therefore crosses the same latitudes of the host during each transit. We calculate WASP-80b to have an impact parameter of 0.2, therefore it is plausible that there are spots at higher latitudes that are not occulted during transit. High latitude spots have been observed in a number of stars (e.g. \citealt{Strassmeier2009}), although early M dwarfs have been observed to show more spots uniformly distributed in longitude and latitude (e.g. \citealt{Barnes2001,Barnes2004}), making the prospect of numerous unocculted spots less likely.

Unocculted spots can induce blueward slopes in transmission spectra \citep{McCullough2014}, mimicking a Rayleigh scattering signature. Unocculted spots cannot be strongly affecting our transmission spectrum as we do not detect a clear scattering signature, and since we do not know whether the star has evenly distributed spots, or has no spots, we have chosen not to adjust the transmission spectra presented here. However, we consider whether they could cause the discrepancy between the transmission spectrum presented here and that of \cite{Sedaghati2017}.

Following the formalism of \cite{McCullough2014}, who considered the effects of unocculted spots on the transmission spectrum of HD\,189733b, it can be shown that the apparent depth $(\tilde{R_P}/\tilde{R_*})^2$, can be related to the measured depth $(R_P/R_*)^2$, the filling factor of spots on the projected stellar surface $f$, and the fluxes of the star and a spot at a particular wavelength, $F_{\lambda}(spot)$ and $F_{\lambda}(star)$, through

\begin{equation}
\left(\frac{\tilde{R_P}}{\tilde{R_*}}\right)^2 = \frac{(R_P/R_*)^2}{1 - f(1-F_{\lambda}(spot)/F_{\lambda}(star))},
\end{equation}

which can be rearranged to find the filling factor, $f$. Using \textsc{atlas9} models of a star with $\mathrm{T_{eff}} = 4000$\,K, [Fe/H] = 0.0, and $\log g = 4.5$, and a spot with a temperature of 3500\,K (as consistent with observations of spots on stars of similar spectral type, \citealt{Berdyugina2005}), we find that the difference in transit depth between our shortest and longest wavelength observations would require a filling factor of 2.6\%. The slope found by \cite{Sedaghati2017} would require unocculted spots with a filling factor of 16\%. If these transmission spectra were due to unocculted spots, then the filling factor of such spots would have to change by at least 13\%. Since photometric modulations of this order are ruled out by the WASP data we conclude that stellar activity is very unlikely to be the cause of the discrepancy between our results and those of \cite{Sedaghati2017}.

\section{Conclusions}

We have presented a ground-based optical transmission spectrum of the hot Jupiter WASP-80b. Our transmission spectrum is best represented by a Rayleigh scattering slope, suggesting the presence of a high altitude haze in the atmosphere of WASP-80b. We see no evidence for the broad wings of the potassium feature as claimed previously by \cite{Sedaghati2017} nor sodium at a resolution of 100\,\AA. Instead, our transmission spectrum is in better agreement with those of \cite{Fukui2014}, \cite{Mancini2014_w80}, \cite{Triaud2015} and \cite{Parviainen2017}. 

Stellar activity is very unlikely to be the cause of the discrepancy between our results and those of \cite{Sedaghati2017} due to WASP-80's lack of photometric modulation \citep{Triaud2013}. Instead, it is possible that there were residual systematics in the VLT/FORS data of \cite{Sedaghati2017}, perhaps related to the LADC, despite the authors' thorough attempts to account for these.

This is the third paper in the LRG-BEASTS programme following the detection of a Rayleigh scattering haze in HAT-P-18b \citep{Kirk2017} and clouds in WASP-52b \citep{Louden2017}. LRG-BEASTS is demonstrating that 4-metre class telescopes can provide transmission spectra with precisions comparable to 8- and 10-metre class telescopes, and are capable of detecting and ruling out model atmospheres.  This work also highlights the importance of independent and repeat studies of hot Jupiters. These studies as part of our larger survey will help to shed light on the prevalence, and physical origins, of clouds and hazes in hot Jupiter atmospheres by increasing the sample of studied hot Jupiters. 

\section*{Acknowledgements}

We thank the anonymous referee for their helpful suggestions and comments which improved the discussion of the manuscript. We also thank Hannu Parviainen and Tom Evans for useful discussions during the preparation of this manuscript. J.K. is supported by a Science and Technology Facilities Council (STFC) studentship. P.W. is supported by an STFC consolidated grant (ST/P000495/1). The reduced light curves presented in this work will be made available at the CDS (http://cdsarc.u-strasbg.fr/). This work made use of the \textsc{astropy} \citep{astropy}, \textsc{numpy} \citep{numpy} and \textsc{matplotlib} \citep{matplotlib} \textsc{python} packages in addition to those cited within the body of the paper. The William Herschel Telescope is operated on the island of La Palma by the Isaac Newton Group in the Spanish Observatorio del Roque de los Muchachos of the Instituto de Astrof\'{i}sica de Canarias. The ACAM spectroscopy was obtained as part of W/2016B/28.

\bibliographystyle{mnras}
\bibliography{wasp80_bib} 

\bsp	
\label{lastpage}
\end{document}